# Observation of Coherent Spin Waves in a Three-Dimensional Artificial Spin Ice Structure

*Sourav Sahoo[1], Andrew May[2], Arjen van Den Berg[2], Amrit Kumar Mondal[1], Sam Ladak[2] and Anjan Barman[1*]*

1. Department of Condensed Matter Physics and Material Sciences, S. N. Bose National Centre for Basic Sciences, Block JD, Sector III, Salt Lake, Kolkata 700 106, India

2. School of Physics and Astronomy, Cardiff University, Cardiff CF24 3AA, UK

*abarman@bose.res.in

Harnessing high-frequency spin dynamics in three-dimensional (3D) nanostructures may lead to paradigm-shifting, next generation devices including high density spintronics and neuromorphic systems. Despite remarkable progress in fabrication, the measurement and interpretation of spin dynamics in complex 3D structures remain exceptionally challenging. Here we take a first step and measure coherent spin waves within a 3D artificial spin ice (ASI) structure using Brillouin light scattering. The 3D-ASI was fabricated by using a combination of two-photon lithography and thermal evaporation. Two spin-wave modes were observed in the experiment whose frequencies showed a monotonic variation with the applied field strength. Numerical simulations qualitatively reproduced the observed modes. The simulated mode profiles revealed the collective nature of the modes extending throughout the complex network of nanowires while showing spatial quantization with varying mode quantization numbers. The study shows a well-defined means to explore high-frequency spin dynamics in complex 3D spintronic and magnonic structures.

Patterned magnetic nanostructures have been studied a great deal during the last few decades due to their interesting spin configurations [1-3] and their potential applications in energy-efficient miniaturized spintronics as well as magnonic devices [4-6] where spin waves (SW) may act as an information carrier. A large volume of work has been done on one-dimensional (1D) and two-dimensional (2D) magnonic crystals (MCs) of different form and geometry [2,7-10]. These include magnetic dot arrays [11], antidot arrays [12], bicomponent magnonic crystal (BMC) [13], nanowires [14] and nanostripes [15]. During the last few years, three-dimensional (3D) nanomagnetism has emerged as a fascinating research field demonstrating novel physical phenomena such as curvature induced anisotropy [16,17], frustration in 3D artificial spin ice (ASI) systems [18,19], 3D magnonic crystals [17,20], noncollinear spin textures such as twisted skyrmion [21], magnetic singularities, e.g. Bloch points [22,23], hopfions [24] and vortex domain walls [25]. On the other hand, 3D magnetic nanostructures have the potential for future applications in magnetic sensors [26], neuromorphic computing [27], ultra-dense data storage devices [28,29] and 2.5D spintronic devices [30]. The main hindrances behind exploration of 3D magnetic nanostructures [30-32] have been their non-trivial fabrication and characterization techniques. The combination of 3D patterning techniques such as focused electron beam-ion deposition (FEBID) [33], two-photon lithography (TPL) [34-37] with sputtered deposition [38], electrodeposition [39] and thermal evaporation [19,40] have emerged as powerful techniques to fabricate 3D complex magnetic nanostructures for investigation of novel phenomena and development of future magnetic devices. Recently, high quality free-standing tetrapod structures have been made from Cobalt nanowires, by utilising TPL and electrodeposition [39]. The SW dynamics from the junction of a tetrapod structure was experimentally measured using a time resolved Kerr microscope [41]. However, the large separation between the tetrapods did not allow the study of coherent magnons in this system. The study of SW dynamics within interconnected 3D magnetic nanostructures is important to first of all build an elementary understanding of SW mode behaviours within such complex systems and subsequently to develop future devices which allow the propagation of spin waves to be controlled in complex 3D circuits. Such structures hold the promise to study coherent magnon states in 3D MCs due to Bragg scattering in all three spatial directions, as well as investigation of anisotropic magnon minibands and Brillouin zone boundaries along high symmetry directions. Theoretical studies of SW dynamics in prototype of 3D interconnected magnetic nanostructures [20,42] have been reported recently. Finally, the realisation of 3D magnetic nanostructures in complex frustrated geometries, such as a 3D-ASI, provides access to a huge number of near degenerate states, providing a platform for reconfigurable magnonic devices [43]. However, the experimental study of coherent SW dynamics in an interconnected 3D-ASI system is currently missing in the literature.

Here, we report upon the experimental measurement of SW modes in a 3D-ASI composed of interconnected nanowires arranged in diamond bond lattice (DBL) structure using conventional Brillouin light scattering (BLS). The 3D-ASI was fabricated by using a combination of TPL and thermal evaporation. Two clear SW modes were observed in the BLS spectra, each of which showed a systematic variation with the applied magnetic field. These experimental results have been understood in the context of 3D micromagnetic simulations, which show the observed modes can be reproduced in the simulation. The simulated mode profiles revealed complex quantized characters with its power distributed over the entire structure.

A 3D array of interconnected nanowires of DBL (3D-DBL) was fabricated by using a three-step process. In the first step, a 3D diamond-bond lattice structure ($50 \times 50 \times 10$ $\mu m^3$) was fabricated upon glass using TPL and subsequent development. In the second step, a layer of gold (30 nm) was deposited upon the sidewalls of the scaffold nanowires. This was achieved by carrying out four separate Au depositions, whereby the sample was mounted at a 30-degree tilt and the in-plane angle was rotated by 90-degrees for each deposition. Finally, a 50-nm-thick $Ni_{81}Fe_{19}$ (Permalloy; Py hereafter) was deposited with the substrate in a flat, zero-tilt position. The deposition of Py on the curved surfaces leads to formation of nanowires with crescent shaped cross section [18]. Overall, the process yields a DBL of Py which is continuous for four layers, in the y-direction, corresponding to a unit cell in thickness [19]

The scanning electron micrographs of the full 3D array is shown in Fig. 1(a) and a magnified view (inset of Fig. 1(a)) shows a constituent tetrapod element of the interconnected nanowire structure. The four sub-lattice layers are annotated in Fig. 1(b). The individual nanowire length is approximately 1000 nm and its width is approximately 260 nm. A deviation up to ±10 nm in the width, and up to ±25 nm in the length of the nanowires was observed. More details of fabrication and characterization of the samples can be found elsewhere [19].

The SW dynamics of the 3D array was measured by using conventional Brillouin light scattering (BLS) technique  [44]. The BLS is a popular tool to measure SW dynamics of magnetic thin films and patterned nanostructures. It is a non-contact and thus non-invasive tool to measure thermally excited SWs without any external excitations under ambient condition. This technique relies upon inelastic scattering of light from the sample due to interaction between incident photons and magnons. A continuous wave of monochromatic laser (wavelength $\lambda = 532$ nm, power = 60 mW) was focused on the sample (Fig. 2(a)) to a spot size of around 40 µm, which is close to the lateral dimensions of the sample. As a result, the SW were measured from almost the entire sample area. The cross polarization between the inelastically backscattered beam and incident beam was exploited to supress the phonon contribution. A Sandercock-type six-pass tandem Fabry–Perot interferometer was used to analyse the frequencies of the scattered beam, in order to extract the SW frequencies. In our experiment, we applied a bias magnetic field ($H$) parallel to the substrate plane as shown in the inset of Fig. 2(a), along a principal axis (x-direction) of the lattice. A schematic of the experimental geometry is shown in Fig. 2(a).

In order to study the SW frequency variation with $H$, the BLS spectra was measured for $k \approx 0$ wave vector in Damon-Eschbach (DE) geometry for different $H$ values varied between 0.6 and 2.0 kOe. Some example BLS spectra from the 3D-ASI are shown in Fig. 2(b). Two intense SW modes were observed in the spectra which are named as M1 and M2. The lower frequency peak (M1) becomes more prominent at larger values of $H$. The higher frequency mode (M2) is quite broad apparently due to unresolved modes and/or inhomogeneous line broadening due to defects and inhomogeneous spin textures in the structure. The BLS spectra were fitted with two-peak Lorentizan functions to extract the SW frequency values. The bias magnetic field variation of SW frequencies is plotted in Fig. 3(a). The SW mode frequency increases monotonically with increasing field values suggesting purely magnetic origin of the modes. Despite the complicated structure of the 3D-ASI and the corresponding demagnetizing factors, we have fitted the most intense mode with Kittel formula, which resulted in a good fit with effective demagnetizing factors at three different axes as presented in section S1 of the Supplementary Information.

To obtain deeper insight into the behaviour of the observed SW modes, we have numerically simulated the SW dynamics in the 3D-ASI system using the GPU-based mumax3 software [45]. A schematic of the simulated diamond-lattice [46] unit cell is shown in Fig. 4 (a) for clarity of understanding our 3D-ASI structure, where the atoms are non-existent and only the bonds are present. A typical simulated static spin configuration of the 3D-ASI structure is shown in Fig. 4(b), which consists of four tetrapod elements, highlighted by different color. The 3D-ASI structure has been designed to match the geometry of a diamond bond lattice, as shown in Fig 4(a). In order to mimic the experimental sample volume, we considered a unit cell of the 3D-ASI in the simulation and applied a 2D periodic boundary condition in the x-z plane, while along the y direction the simulated structure contains four layers similar to the experimental sample. The simulated unit cell of the 3D-ASI consists of crescent shaped nanowires with dimensions similar to the experimental sample. The sample was divided into cuboidal cells of size $5 \times 5 \times 5$ nm$^3$. The cell size is taken below the exchange length of Py ($\approx$ 5.2 nm). The material parameters used in the simulation are: gyromagnetic ratio $\gamma = 17.6$ MHz/Oe, saturation magnetization $M_s = 860$ emu/cc, anisotropy field $H_K = 0$ and exchange stiffness constant $A_{ex} = 13 \times 10^{-6}$ erg/cm for Py [47]. The equilibrium magnetic states were obtained by relaxing the sample under study at a fixed bias magnetic field along x direction as defined in Fig. 4(c). The magnetization configuration of the equilibrium state ($m_x$ component) at $H = 1.6$ kOe is shown in Fig. 4(c) which shows a saturated state along x direction with negligible demagnetize region. The equilibrium state magnetization configuration of other two components ($m_y$ and $m_z$) are shown in Fig. S2 of the Supplementary Information. For simulation of the SW dynamics, a square shaped pulsed magnetic field with peak amplitude of 20 Oe along y direction with rise and fall time of both 10 ps and duration of 20 ps, was applied to the equilibrium magnetic state. The SW spectra were calculated by taking FFT of the simulated time-domain magnetization ($m_y$ component). Some typical simulated SW spectra are shown in Fig. 3 (b). The simulated spectra show additional peaks which were either not resolved due to the line broadening or insufficient sensitivity due to their smaller power in the experiment. Nevertheless, both the experimental peaks could be identified in the simulation and the variation of their frequencies with bias magnetic field is plotted as dotted lines in Fig. 3(a). A qualitative agreement between the experimental and simulated bias-field variation of SW frequencies is found. To understand the spatial nature of the coherent SW modes in the 3D-ASI, we have further analyzed the simulated data from Mumax$^3$ using a home-built post processing code [48]. The scheme of the post processing is briefly discussed in section S3 of the Supplementary Information. The mode profiles were calculated along different planes at positions 1 and 2, as shown in Fig. 4(c). The coordinate system in Fig. 4(c) is defined such that the projections of top and bottom layers on the x-z plane become parallel to x′ and z′ axis, respectively. The simulated powermaps of the experimentally observed SW modes are shown in Fig. 5 for $H = 1.6$ kOe. The powermaps of additional modes present only in the simulated spectra are presented in the Supplementary Information. The powermaps calculated along x′-y′ plane at position 1, is shown in Fig. 5(a). It captures the lateral view of the top layer and cross-sectional view of one nanowire of the bottom layer. The power of M1 is localized at the junction of each bipod but the power of M2 is extended throughout the 3D-ASI nanowire network. The phase of these modes revealed quantized nature with quantization number $n = 7$ and 5 for M1 and M2, respectively. The cross-sectional view of the nanowire in the bottom layer revealed M1 to have quantized nature ($n = 2$) with low power, while M2 possess uniform behaviour with reasonably high power. To explore the mode profiles further, we took another slice on y′-z′ plane at position 2 (Fig. 5(b)). Here, power is distributed over the whole nanowire network for both modes. M1 again shows quantized nature with $n = 9$, while M2 also shows quantized nature with $n = 7$. The mode profile behaviour at the cross section of nanowires was analysed by taking slice along x′-y′ plane at position 1, as shown in Fig. 5(c). Here, M1 forms a quantized mode with $n = 4$ at the cross section of the junction, while M2 also forms a quantized mode with $n = 2$ at the cross sections. The overall power profile suggests that M1 is primarily localized at the nanowire junctions while M2 is extended over the nanowire networks. The phase profiles revealed the 3D nature of these modes with different quantization along the lateral direction of nanowires and at the cross sections. We have applied the magnetic field along a principal axis (along x-axis, Fig. 4(a)) which is an in-plane symmetry axis for the 3D-ASI structure. However, the crescent cross section of the nanowires and the 3D geometry of ASI may create a 3D varying magnetic potential within this nanostructure leading to the quantized nature for both SW modes with different quantization number.

In summary, we have exploited a novel method to fabricate a complex 3D-ASI structure of interconnected nanowires arranged in diamond bond-like lattice using TPL and thermal evaporation process. We have studied thermal magnons in this 3D-ASI using BLS spectroscopy which revealed two clear SW modes. The SW mode frequencies show good stability and monotonic variation over a broad range of bias magnetic field. We have performed 3D micromagnetic simulations to reproduce the SW spectra and obtained insights into the spatial nature of the SW modes in this 3D-ASI. The SW modes exhibit different spatial characters from localized to extended nature having different mode quantization numbers. Some additional modes in the simulation were either not resolved or detected in the experiment, which called for more precise experiment to detect those. On the other hand, fabrication of even higher quality samples extended equally in all three dimensions will be helpful to understand the SW propagation along all high-symmetry axes in these structures. To this end, optical exploration of SW modes in this interconnected 3D structure will open the pathway for exciting new possibilities of 3D magnonic devices.

**Acknowledgements:** AB gratefully acknowledges the financial support from S. N. Bose National Centre for Basic Sciences, India (Grant No. SNB/AB/18-19/211). SL gratefully acknowledges funding form the Engineering and Physics Research Council (EP/R009147/1). SS and AKM acknowledge S. N. Bose National Centre for Basic Sciences for senior research fellowship

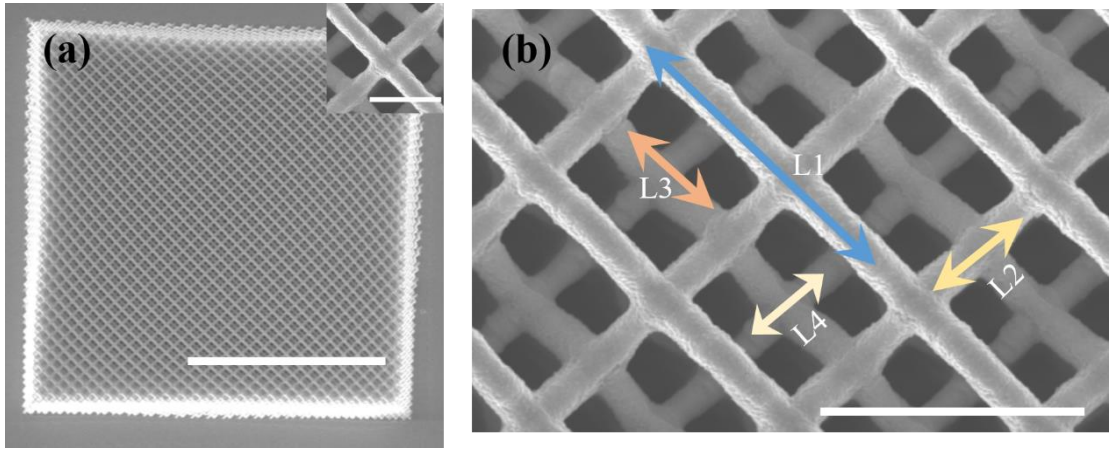

Figure 1: (a) Scanning electron micrograph showing top view of full array of 3D-ASI with size 50×50×10 µm³, the scale bar is 25 µm. A constituent tetrapod element is shown in the inset by capturing a zoomed view of 3D array where the scale bar is 1 µm (b) A magnified view of the interconnected nanowires in the lattice is shown. Four sub-lattice layers are highlighted. Scale bar in (b) is 2 µm.

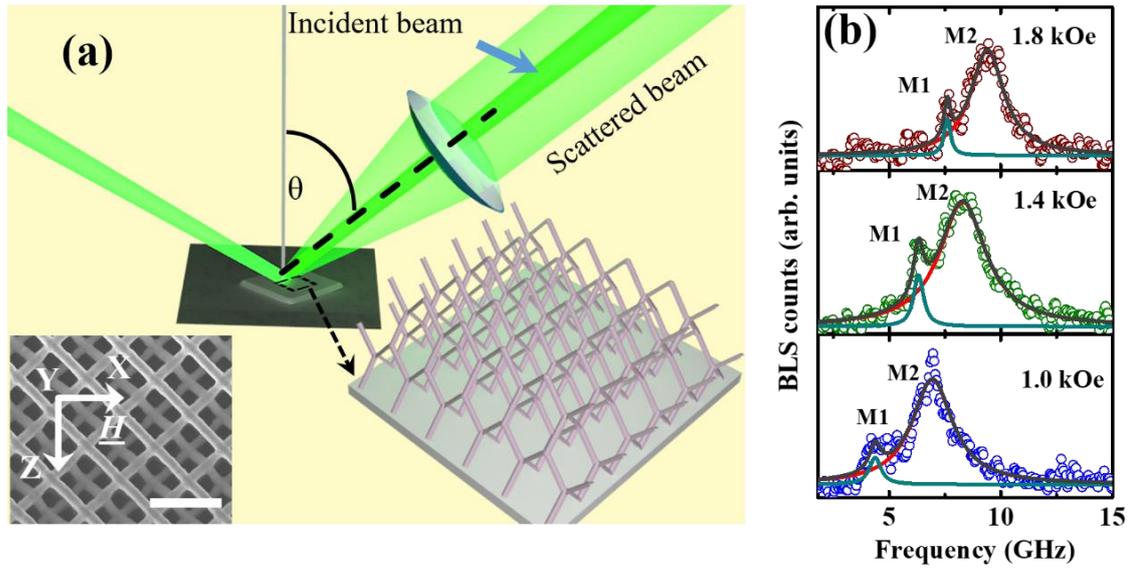

Figure 2: (a) Schematic of measurement geometry. The measurement was performed at θ = 0°. The applied field (*H*) direction is shown at the left side of the inset and the scale bar is 2µm. The 3D-ASI network is presented by a schematic on the right of the image. (b) BLS spectra for three different field values are shown and the field stated are in units of kOe. Open circles present the experimental data points. Here, cyan and red color solid lines present the fitting of individual peak and the grey color solid line present the resultant of the multi-peak fitting.

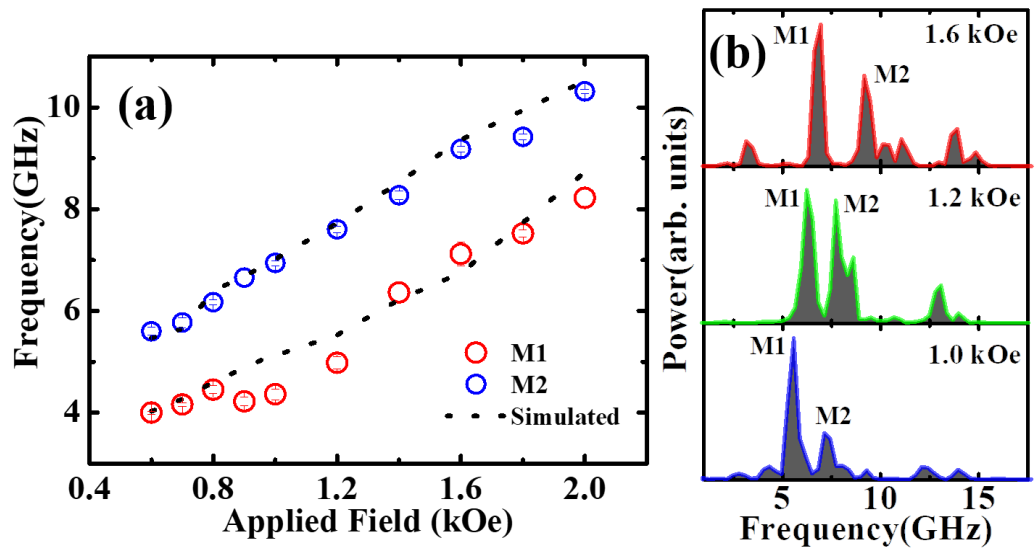

Figure 3: (a) Spin-wave frequencies of mode 1 (M1) and mode 2 (M2) are plotted as a function of applied magnetic field. Here, symbols present the experimentally measured data points. (b) Simulated spin-wave spectra for three intermediate field values are shown. The dotted lines in (a) present simulated results.

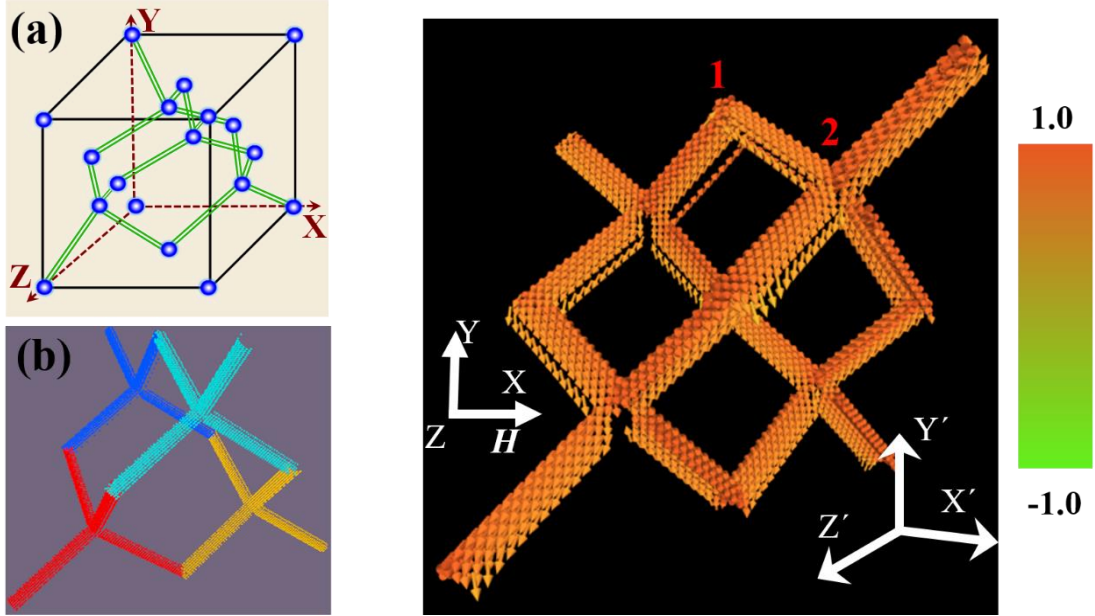

Figure 4: (a) Schematic of diamond-lattice unit cell. Here, the green color double lines present the bond position and blue spheres present the atom position. The axes are defined by a dashed line. (b) A representative simulated unit cell of 3D-ASI is shown. One unit cell consists of four tetrapod elements, each of them are highlighted by different color for better visualization. A one-to-one correspondence between the diamond bond lattice and 3D-ASI (omitting the atoms) could be found from (a) and (b). (c) Magnetization configuration of the equilibrium state ($m_x$ component) at $H$ = 1.6 kOe (along x direction; in-plane) is shown. A new coordinate system (x´y´z´) is defined and two positions are marked as 1 and 2 for mode profile analyses.

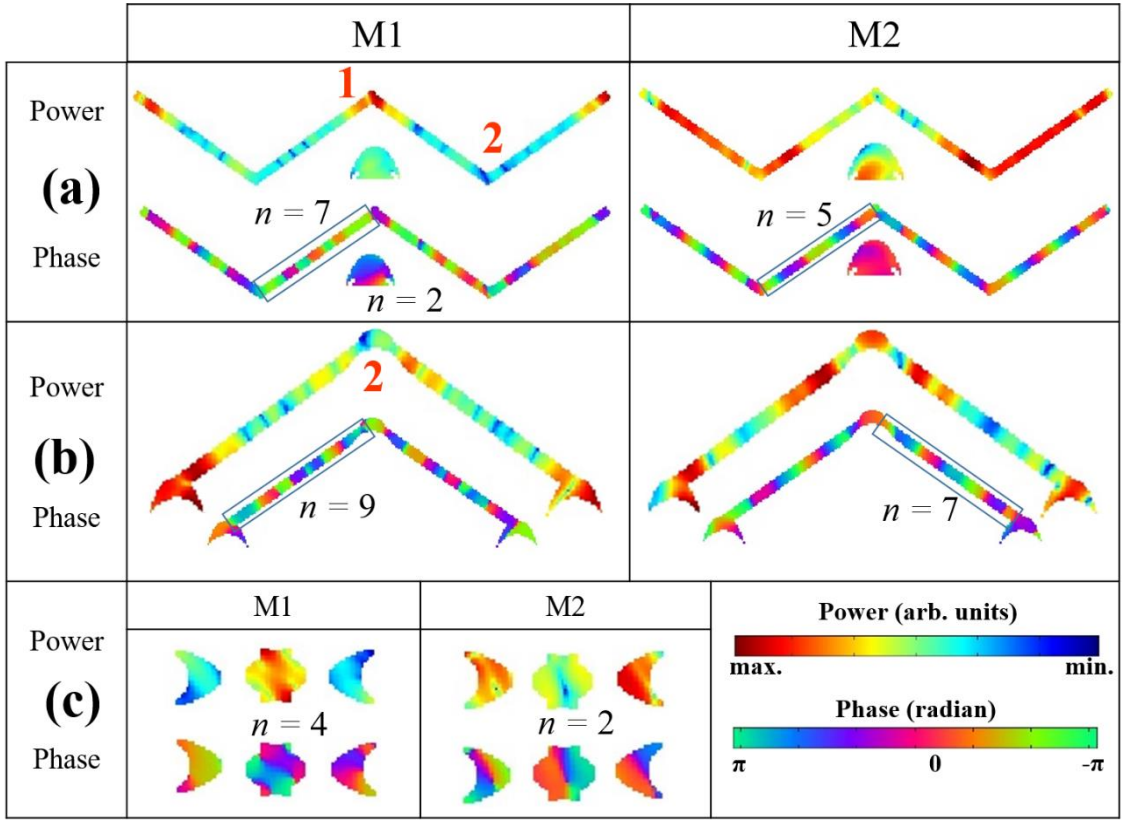

Figure 5: Spin-wave mode profiles are calculated at two different positions of the 3D-ASI structure for $H = 1.6$ kOe by taking slice along - (a) x´-y´ plane at position 1, (b) y´-z´ plane at position 2 and (c) x´-z´ plane at position 1. The x´y´z´ coordinate system and positions (1 and 2) are shown in figure 4(c). The power profiles are shown in upper part and corresponding phase profiles in lower part of each panels. The color bars are shown at bottom corner. The numbers in red color show the slice taking position, as shown figure 4(c).

# Supplementary information

## Observation of Coherent Spin Waves in a Three-Dimensional Artificial Spin Ice Structure

*Sourav Sahoo[1], Andrew May[2], Arjen van Den Berg[2], Amrit Kumar Mondal[1], Sam Ladak[2] and Anjan Barman[1\*]*

1. Department of Condensed Matter Physics and Material Sciences, S. N. Bose National Centre for Basic Sciences, Block JD, Sector III, Salt Lake, Kolkata 700 106, India

2. School of Physics and Astronomy, Cardiff University, Cardiff CF24 3AA, UK

*abarman@bose.res.in

## S1. Bias Field Dependent Spin-Wave Frequency:

Despite the complex structure of the 3D-ASI studied in this work, we attempted to fit the bias field dependent frequency of the dominant spin-wave mode of the studied structure with Kittel formula [1]. Considering the applied bias magnetic field along x direction and zero magneto-crystalline anisotropy in our sample we write the Kittel formula as:

$$f = \frac{\gamma}{2\pi}[(H + H_{d1})(H + H_{d2})]^{\frac{1}{2}} \qquad [S1]$$

Here, $\gamma$ is gyromagnetic ratio, $H$ is external applied magnetic field, $H_{d1} = (N_z - N_x)M_s$ and $H_{d2} = (N_y - N_x)M_s$ are the effective demagnetizing fields originating from the shape of the studied structure, and $M_s$ is saturation magnetization. Here, $N_x$, $N_y$ and $N_z$ are the demagnetizing factors along x, y and z direction, respectively. The fitted curve is shown in Fig. S1. The experimental data was replotted from Fig. 3(a) of the article. From the fit we find the values of demagnetizing factor as $N_x = 2.54$, $N_z = 2.82$ and $N_y = 7.2$ considering the values of $\gamma = 17.6$ MHz/Oe and $M_s = 860$ emu/cc. The extracted values of demagnetizing factors suggest that our sample has a relatively high demagnetization contribution along out-of-plane direction (along y, shown in Fig. S2).

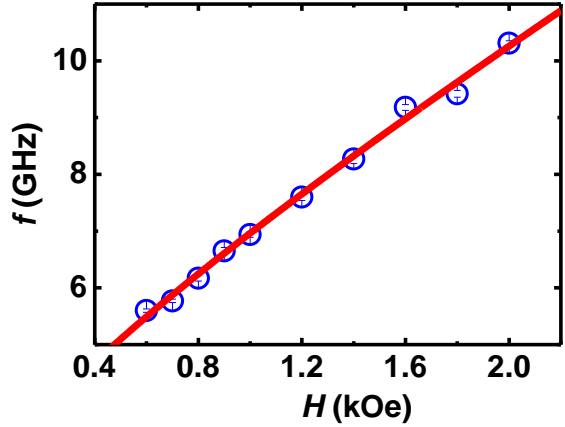

Figure S1: Experimental data of bias field dependent spin-wave frequency (open symbols) fitted with Kittel formula of Eq. S1 (solid line).

## S2. Static Magnetization Configuration of the Sample:

The equilibrium state (static) magnetization configuration of two orthogonal components ($m_y$ and $m_z$) at $H = 1.6$ kOe applied along x direction ($m_x$ is already shown in Fig. 4(c) of the article) is presented in Fig. S2. Two orthogonal components show prominent demagnetized regions in the magnetization configuration. A distinct difference in the spin structures in the out-of-plane component ($m_y$ component), including the demagnetization regions between intra- and inter-nanowires branches is observed. The out-of-plane configuration ($m_y$) shows more dominant demagnetized state compared to the in-plane configuration ($m_z$) which also validates the extracted results of higher demagnetizing factor ($N_y$) along the out-of-plane direction as presented in **S1**.

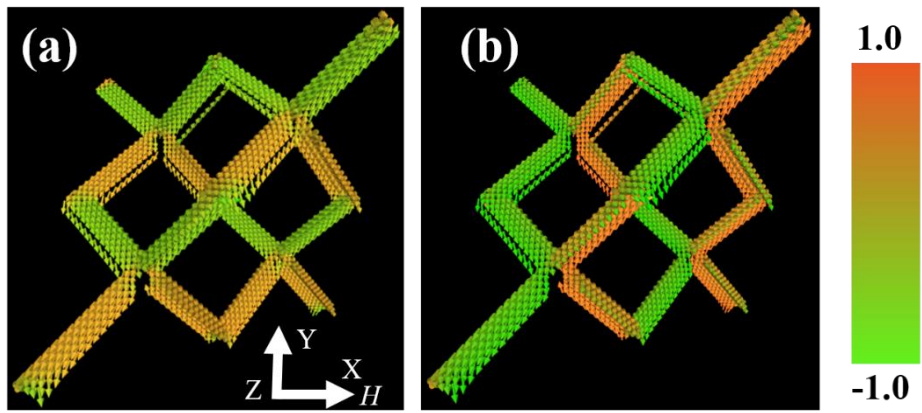

Figure S2: Static magnetization configuration of (a) $m_z$ component and (b) $m_y$ component of the 3D-ASI sample at $H = 1.6$ kOe applied in the plane of the sample (along x direction).

**S3**. Spin-Wave Mode Profile Calculation Procedure:

In the dynamic simulations mumax3 software generates ".ovf" files with spatial distribution of magnetization ($M(t,x,y,z)$) at a particular given time which contains the information of superposed multiple resonant modes. The calculation of power and phase profiles of the multiple resonant modes requires further processing of the data. In order to extract the power and phase profiles of the multiple resonant modes we analysed the ".ovf" files using our home built post processing code DOTMAG [2,3]. To calculate the power and phase profiles in the frequency domain, it takes the fast Fourier transformation (FFT) of time-domain data along a plane of the sample by keeping one of the coordinates fixed (either x or y or z). If we fix $z = z_1$ then the FFT is taken along the x-y plane: $\tilde{M}^{z_1}(f,x,y) = FFT\left(M^{z_1}(t,x,y)\right)$. Then the power and phase is calculated for a resonant mode of particular frequency $f = f_1$ which can be written as:

Power: $\qquad P^{z_1 f_1}(x,y) = 20 log_{10}\left|\tilde{M}^{z_1}(f_1,x,y)\right|$ and

Phase: $\qquad \phi^{z_1 f_1}(x,y) = \text{atan2}\left(Im\left(\tilde{M}^{z_1}(f_1,x,y)\right), Re\left(\tilde{M}^{z_1}(f_1,x,y)\right)\right)$

**S4**. Calculated Spin-Wave Mode Profiles of Additional Modes Observed Only in the Simulation:

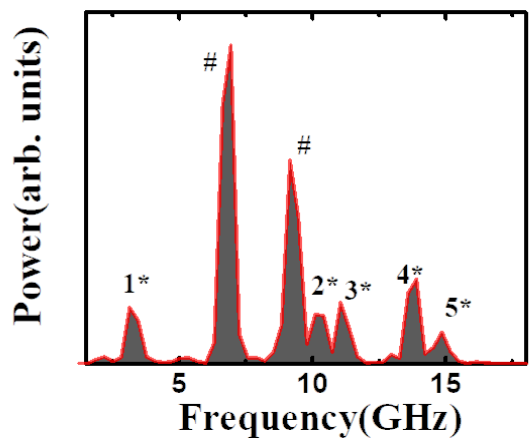

Figure S3: Simulated spin-wave spectra at $H = 1.6$ kOe. The experimentally observed modes are marked by '#' while the additional modes observed explicitly in simulation are numbered as 'digit*' from lower to higher frequencies.

In the simulated SW spectra, we have observed some additional modes, which were either not resolved or detected due to limitations in resolution and or detection sensitivity. However, these modes are fundamentally important, as they are characteristic modes of this structure. Hence, we have simulated mode profiles of those modes too for the completeness of study. The modes are numbered from lower to higher frequencies (Fig. S3). Figure S3 is repeated from Fig. 3 (b) of the article for ready correlation with the simulated mode profiles. The SW mode profiles are presented in Fig. S4. Here, M1* shows quantized nature with quantization number $n = 5$ (Fig. S4 (a)) and 11 (Fig. S4 (b)). The cross-sectional view shows that it forms quantized mode across the nanowire (Fig. S4 (a)) and at the junction (Fig. S4 (c)) with $n = 2$ and $n = 3$, respectively. Here, M2* and M3* show quantized behaviour with power distributed over the nanowires (Fig. S4 (a) and (b)). The quantization number is found to be $n = 6$ and 7 for M2* and $n = 8$ and 6 for M3, as shown in Fig. S4 (a) and (b). The cross-sectional view at the junction (Fig. S4 (c)) shows the quantized nature of the modes with $n = 3$ and $n = 2$ for M2* and M3*, respectively. Here, two higher frequency modes also show the quantized nature along the lateral directions of connected nanowires (Fig. S4 (d) and (e)) with $n = 9$ and 12 for M4* and $n = 11$ and 15 for M5*. The cross-sectional view of the nanowire (Fig. S4 (d)) shows mixed quantized nature of the modes (M4* and M5*) with quantization number 2 each along two mutually orthogonal direction. The cross-sectional view at the junction (Fig. S4 (f)) shows the quantized nature with $n = 2$ and $n = 3$ for M4* and M5*, respectively. The analyses of SW mode profiles reveal the quantized nature of these observed additional modes in simulated spectra with significant amount of power distributed all over the nanowires. The phase profiles reveal the 3D nature of these modes with different quantization number and nature along the different lateral directions and cross sections of the connected nanowires of the 3D ASI structure.

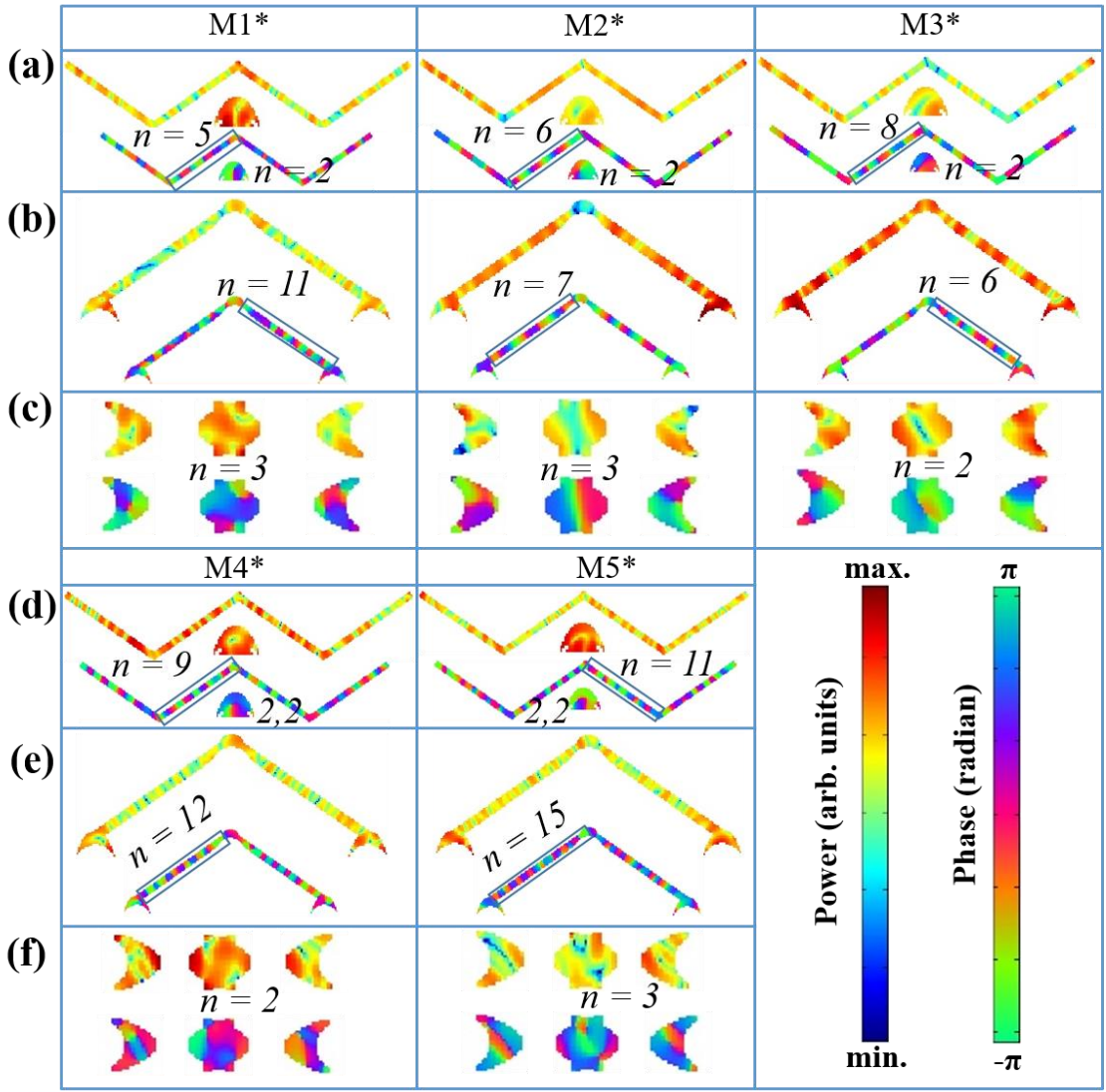

Figure S4: Spin-wave mode profiles calculated at different points of 3D-ASI structure at $H = 1.6$ kOe by taking slice along (a), (d) x´-y´ plane at point 1, (b), (e) y´-z´ plane at point 2 and (c), (f) x´-z´ plane at point 1. The mode profiles of M1, M2 and M3 are shown in (a), (b) and (c), while (d), (e) and (f) show the mode profiles of M4* and M5*. The x´-y-´z´ coordinate and positions (1 and 2) are presented in Fig. 4(c) of the article. The power profiles are shown in upper part and corresponding phase profiles in lower part of each panel. The color bars are presented at bottom corner.

## Supplementary References: